\documentclass[aps,pr,reprint,superscriptaddress]{revtex4-1}
\usepackage{graphicx}
\usepackage{dcolumn}
\usepackage{bm}
\usepackage{color}
\usepackage{xcolor}
\usepackage{multirow}
\usepackage{hyperref}
\usepackage{amssymb}
\usepackage{color}
\usepackage{graphics}\usepackage{epsfig}
\usepackage{footnote}
\usepackage{amsmath}
\usepackage{multirow}
\usepackage{graphicx}
\usepackage{tabularx}
\usepackage{physics}
\usepackage[normalem]{ulem}
\graphicspath{{Figs/}}
\setcitestyle{square,numbers} 
 
\newcommand{\be}{\begin{equation}}
\newcommand{\ee}{\end{equation}}
\newcommand{\bea}{\begin{eqnarray}}
\newcommand{\eea}{\end{eqnarray}}
\renewcommand{\vec}[1]{\boldsymbol{#1}}

\newcommand{\bt}{\begin{table}}
\newcommand{\et}{\end{table}}

\begin{document}
\title{Real-space obstruction control of the shift current in two-dimensional Quantum Spin Hall insulators}
\author{Michele Bagaglini}
\email{michele.bagaglini@graduate.univaq.it}
\affiliation{Department of Physical and Chemical Sciences, University of L'Aquila, Via Vetoio 10, I-67100 L'Aquila (Italy)}
\affiliation{CNR-SPIN c/o Department of Physical and Chemical Sciences, University of L'Aquila, Via Vetoio 10, I-67100 L'Aquila (Italy)}
\author{Mattia Iannetti}
\email{mattiaiannetti@gmail.com}
\affiliation{Department of Physical and Chemical Sciences, University of L'Aquila, Via Vetoio 10, I-67100 L'Aquila (Italy)}
\affiliation{CNR-SPIN c/o Department of Physical and Chemical Sciences, University of L'Aquila, Via Vetoio 10, I-67100 L'Aquila (Italy)}
\affiliation{Institut f\"ur Theoretische Physik und Astrophysik, Universit\"at W\"urzburg, D-97074 W\"urzburg, Germany}
\affiliation{W\"urzburg-Dresden Cluster of Excellence ctd.qmat, Universit\"at W\"urzburg, D-97074 W\"urzburg, Germany}
\author{Jonas Erhardt}
\affiliation{Physikalisches Institut, Universit\"at W\"urzburg, D-97074 W\"urzburg, Germany}
\affiliation{W\"urzburg-Dresden Cluster of Excellence ctd.qmat, Universit\"at W\"urzburg, D-97074 W\"urzburg, Germany}
\author{J\"org Sch\"afer}
\affiliation{Physikalisches Institut, Universit\"at W\"urzburg, D-97074 W\"urzburg, Germany}
\affiliation{W\"urzburg-Dresden Cluster of Excellence ctd.qmat, Universit\"at W\"urzburg, D-97074 W\"urzburg, Germany}
\author{Ralph Claessen}
\affiliation{Physikalisches Institut, Universit\"at W\"urzburg, D-97074 W\"urzburg, Germany}
\affiliation{W\"urzburg-Dresden Cluster of Excellence ctd.qmat, Universit\"at W\"urzburg, D-97074 W\"urzburg, Germany}
\author{Giorgio Sangiovanni}
\affiliation{Institut f\"ur Theoretische Physik und Astrophysik, Universit\"at W\"urzburg, D-97074 W\"urzburg, Germany}
\affiliation{W\"urzburg-Dresden Cluster of Excellence ctd.qmat, Universit\"at W\"urzburg, D-97074 W\"urzburg, Germany}
\author{Gianni Profeta}
\affiliation{Department of Physical and Chemical Sciences, University of L'Aquila, Via Vetoio 10, I-67100 L'Aquila (Italy)}
\affiliation{CNR-SPIN c/o Department of Physical and Chemical Sciences, University of L'Aquila, Via Vetoio 10, I-67100 L'Aquila (Italy)}

\begin{abstract}

The search for materials exhibiting a large bulk photovoltaic effect has recently expanded to low-band-gap systems capable of operating in the infrared regime, with potential applications in optoelectronics. Among the microscopic mechanisms contributing to this phenomenon, the shift current is one of the dominant sources of photocurrent generation. Owing to its geometric origin, the shift current is not only determined by the optical transition probability but also by the real-space displacement associated with the optical excitation, making non-centrosymmetric materials with obstructed electronic wave functions particularly promising candidates. In this work, using a tight-binding model of a quantum spin Hall insulator on a triangular lattice, we investigate the nonlinear photocurrent across trivial, quantum spin Hall, and higher-order topological insulating phases. We show that obstructed topological phases exhibit a substantially enhanced shift-current response, despite the similar electronic band structure. We identify the origin of this enhancement in optical transitions between states belonging to distinct Wannier-center sectors, which allow for a large real-space displacement upon photoexcitation and, depending on the topological phase, control the sign of the shift current. Our results extend the search for controllable bulk photovoltaic effects to materials characterized by real-space electronic obstruction and identify indenene, a monolayer of In on SiC(0001), as a promising candidate for realizing this phenomenon.

\end{abstract}

\maketitle

\section{Introduction}

In crystalline materials, the electronic response is determined by both their energy spectrum and the geometry of their quantum states. This distinction is particularly consequential in nonlinear optics, where the response can probe information of the wavefunction. A prominent example is the shift current (SC): a direct current generated in a non-centrosymmetric crystal under illumination by linearly polarized light. Unlike conventional photocurrents driven by carrier diffusion or built-in electric fields, the SC originates from a coherent displacement of electronic charge during an interband optical transition~\citep{tan2016shift,cook2017design,tan2016shift,spanier2016power,sipe2000second}. Its magnitude therefore depends both on the probability of the transition and on the shift vector, a gauge-invariant geometrical quantity that describes the associated real-space displacement. Previous studies have established that nonlinear photocurrents can be enhanced by band inversions, quantum-geometric singularities and external control parameters such as strain or electric fields~\citep{sauer2023shift,esteve2025excitons,Inacio2026_bias,Yang2024_strain,Kaner2020_strain,Bagaglini_moire}. However, it remains unclear whether a topological distinction formulated directly in real space can provide a systematic principle for designing large nonlinear photocurrents.\\
Obstructed topological phases offer the possibility to address this question. In these systems, crystalline symmetry constrains the electronic charge centers to positions that differ from the ionic sites. Distinct insulating phases can possess similar dispersions while differing in the spatial organization of their electronic wavefunctions~\citep{Bauernfeind_2021_indenene_topology,Eck_2022_indenene_topology,Erhardt_2022_indenene_topology,Eck_2023_indenene_high_order_topology,eck_2023_thesis_topology_indenene}. Because an optical transition between states centered at inequivalent real-space positions necessarily transfers electronic charge within the unit cell, real-space obstruction provides a possible microscopic source of large shift vectors. This connection could turn the location of the electronic charge centers into a design variable for nonlinear optoelectronic responses as the obstruction strictly depends on the topological parameters, possibly tunable by external control nobs.\\
Here we show that real-space obstruction enhances the SC in quantum spin Hall insulators (QSHI) on a triangular lattice. Across the phase diagram of a multiorbital model, we compare a conventional insulator with two QSHI phases and a higher-order topological phase (HOTI). We find that the obstructed phases support substantially larger nonlinear photocurrents even when their available optical phase space, as measured by the joint density of states (JDOS), remains comparable to that of the trivial phase. By resolving the optical response into transitions between distinct Wannier-center sectors, we trace this enhancement to a larger real-space displacement of the photoexcited charge. The redistribution of the real-space character of the electronic states across the topological phase diagram is accompanied by pronounced changes in both the magnitude and the spectral profile of the SC, including sign differences between distinct insulating phases. Our analysis suggest a general framework in which the SC response can be modified by controlling the driving topological parameters. \\
Finally, we identify indenene, a monolayer of indium deposited on SiC(0001), as a realistic material platform where this mechanism can be realized~\citep{Bauernfeind_2021_indenene_topology,Erhardt_2022_indenene_topology,Erhardt_2025_indenene_topology,Schmitt_2024_indenene_topology_QSHI,Eck_2023_indenene_high_order_topology,Schmitt_2024_indenene_topology_OPTICAL}. Indenene shows a QSHI gap driven by the substrate-induced out-of-plane inversion-symmetry breaking, and is experimentally established as a real-space obstructed material~\citep{Bauernfeind_2021_indenene_topology}. Furthermore, graphene capping has been proposed as a viable strategy to stabilize indenene under ambient conditions~\citep{Erhardt_2024_indenene_topology}. These properties make indenene a candidate not only for observing topology-driven nonlinear optical responses, but also as a system for low-energy devices~\citep{Song2026_THz,Ogino2024_THz,Kaplan2022_THz,Okamura2026_THz}, e. g. for terahertz shift-current applications.

\section{Hamiltonian of Quantum Spin-Hall insulators on a triangular lattice}

\begin{figure*}
\centering
\includegraphics[width=2\columnwidth]{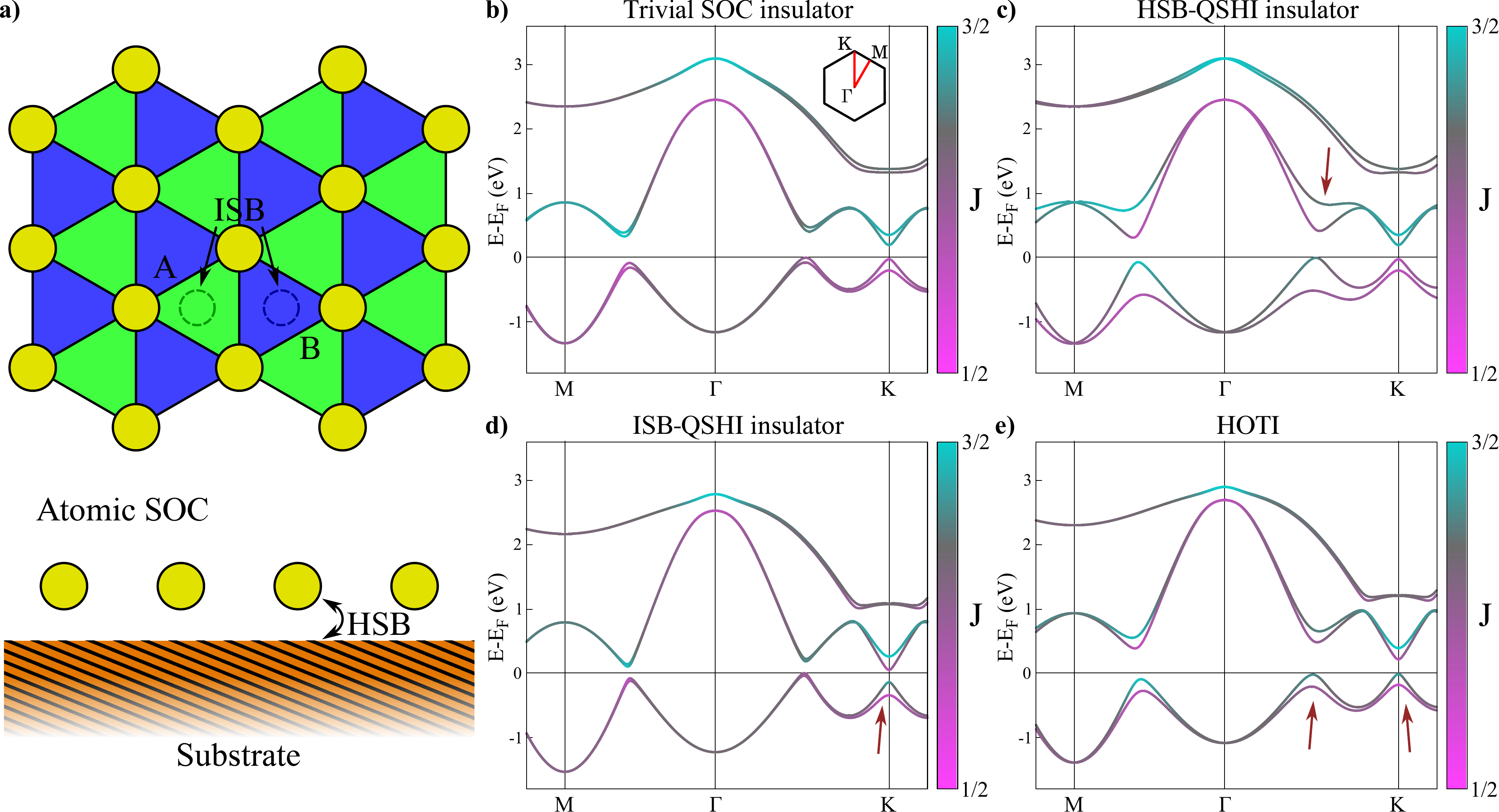}
\caption{\textbf{a}) Sketch of the triangular lattice on a substrate with C$_{3v}$-symmetric. The p-shell atoms are shown in yellow, the in-plane inversion symmetry breaking of the unit cell is marked by blue and green triangles and given by the inequivalence of A and B sites. The out-of-plane inversion symmetry breaking term is given by the substrate while the p-shell atoms provide the spin-orbit-coupling. \textbf{b-e} band structures of $H$ obtained with parameters reported in Table~\ref{table:TBparameters} along the $M-\Gamma-K$ path of the Brillouin zone as shown in the inset. The color code indicates the expectation value of the total angular momentum $J$. Red arrows highlight band inversions happening in $\Gamma-K$ direction with respect to the trivial insulating phase.
}
\label{fig:structure}
\end{figure*}

A simple yet versatile model capable of hosting multiple topological phases and real-space obstruction is the quantum spin Hall insulator (QSHI) on a triangular lattice, originally introduced in Refs.~\cite{Eck_2022_indenene_topology,Eck_2023_indenene_high_order_topology,eck_2023_thesis_topology_indenene,Bauernfeind_2021_indenene_topology,Erhardt_2025_indenene_topology}. The model describes $p$-shell atoms deposited on a substrate with two electrons per site and hopping amplitudes parametrized within the Slater-Koster (SK) formalism. In addition to the hopping processes on the triangular lattice, the electronic structure is shaped by two inversion-symmetry-breaking mechanisms, namely an out-of-plane (horizontal) inversion-symmetry-breaking term (HSB) and an in-plane inversion-symmetry-breaking term (ISB), both induced by the substrate, together with an atomic spin-orbit coupling (SOC), as illustrated in Fig.~\ref{fig:structure}a).
Following Refs.~\cite{Eck_2022_indenene_topology,Eck_2023_indenene_high_order_topology,eck_2023_thesis_topology_indenene,Bauernfeind_2021_indenene_topology,Erhardt_2025_indenene_topology}, the Hamiltonian can be written as

\begin{equation}
\label{Eq:TB}
\hat{H}=
\hat{H}_0+
\hat{H}_{\rm HSB}+
\hat{H}_{\rm ISB}+
\hat{H}_{\rm SOC},
\end{equation}

where $\hat{H}_0$ describes the "free-standing" lattice, $\hat{H}_{\rm HSB}$ and $\hat{H}_{\rm ISB}$ denote the HSB and ISB terms, respectively, and $\hat{H}_{\rm SOC}$ accounts for the atomic SOC. $\hat H_0$ is written in terms of the $p_x$, $p_y$, and $p_z$ orbitals located at the origin of the unit cell and given by the SK parametrization~\citep{Slater_1954_SK_TB}. The HSB term models the presence of the substrate and mixes the $p_z$ and $p_{x,y}$ orbitals with a strength given by $\lambda_{\rm HSB}$~\citep{Eck_2022_indenene_topology,Erhardt_2025_indenene_topology}. The ISB is introduced through an auxiliary dispersionless $s$ orbital located at $(2/3,1/3)$ in terms of the lattice vectors, incorporating the ISB without the use of a down-folding procedure mapping the $s$-subspace on the $p$ shell \cite{eck_2023_thesis_topology_indenene,Eck_2022_indenene_topology,Erhardt_2025_indenene_topology}. Finally, the atomic SOC contribution is encoded in $\hat{H}_{\rm SOC}$ with strength $\lambda_{SOC}$.\\
This TB model (see~\citep{SM} for a complete derivation) exhibits a rich topological phase diagram, as discussed in Refs.~\cite{Eck_2022_indenene_topology,eck_2023_thesis_topology_indenene,Erhardt_2022_indenene_topology,Erhardt_2024_indenene_topology,Erhardt_2025_indenene_topology,Bauernfeind_2021_indenene_topology,Eck_2023_indenene_high_order_topology}, and hosts insulating phases that cannot be adiabatically connected to an atomic limit without relocating the electronic Wannier centers away from the ionic positions. Such a real-space obstruction generates electronic states with distinct spatial distributions, providing the microscopic ingredients for large shift vectors and enhanced nonlinear photocurrents.\\
To this end, we consider a next nearest-neighbor representation of the Hamiltonian $\hat{H}$. The non-vanishing SK parameters are fixed to $V^{\sigma}=0.7~{\rm eV}$, $V^{\pi}=-0.15~{\rm eV}$, and $V^{\pi}_z=-0.25~{\rm eV}$, while the onsite energy of the $p_z$ orbital is set to $E_z=-0.7~{\rm eV}$ with respect to the $p_{x/y}$ orbitals. The onsite energy of the auxiliary $s$ orbital is chosen as $E_s=4|E_z|$. The phase diagram of $\hat{H}$ as a function of $\lambda_{\rm ISB}$, $\lambda_{\rm HSB}$, and $\lambda_{\rm SOC}$ can be summarized by the representative parameter reported in Table~\ref{table:TBparameters}, which identify four insulating phases: a trivial SOC insulator with no obstruction, an obstructed HSB-driven QSHI, an obstructed ISB-driven QSHI, and an obstructed higher-order topological insulator (HOTI).\\
The corresponding band structures are shown in Fig.~\ref{fig:structure}b-e, where the color scale denotes the expectation value of the total angular momentum $J=L+S$. The evolution across the phase diagram is characterized by band inversions, highlighted by red arrows, which signal a redistribution of the orbital and spatial character of the Bloch states. As shown in Refs.\cite{Bauernfeind_2021_indenene_topology,Eck_2022_indenene_topology,eck_2023_thesis_topology_indenene,Eck_2023_indenene_high_order_topology,Schmitt_2024_indenene_topology_QSHI}, this reorganization of the electronic wave functions is accompanied by a real-space obstruction between occupied and unoccupied states. Interestingly, the arrangement of the Wannier centers across the different topological phases allows for sign reversals of the SC. In particular, when the energetic ordering of the Wannier-center sectors is inverted, the direction of the real-space displacement changes, leading to a flip of the shift-current. Experimentally, this can be controlled, for example, by forcing the growth of only one termination of the C$_{3v}$ substrate considered in this work~\citep{Erhardt_2025_indenene_topology,Erhardt_2024_indenene_topology}.

\begin{table*}[t]
    \centering
    \begin{tabular}{c|c|c|c|c}
         &  SOC-driven trivial insulator & HSB-driven QSHI & ISB-driven QSHI & HOTI \\
    \hline
    $\lambda_{\rm SOC}$  (eV)&    0.6    &       0.6       &       0.25      &  0.2 \\
    \hline
    $\lambda_{\rm ISB}$  (eV)&    0.4    &       0.4       &       0.6       &  0.6 \\
    \hline
    $\lambda_{\rm HSB}$  (eV)&    0.02   &       0.15      &       0.02      &  0.12
    \end{tabular}
    \caption{Tight-binding parameters representative of different topological phases.}
    \label{table:TBparameters}
\end{table*}

\section{Shift Current calculation}

\begin{figure}
    \centering
    \includegraphics[width=1\linewidth]{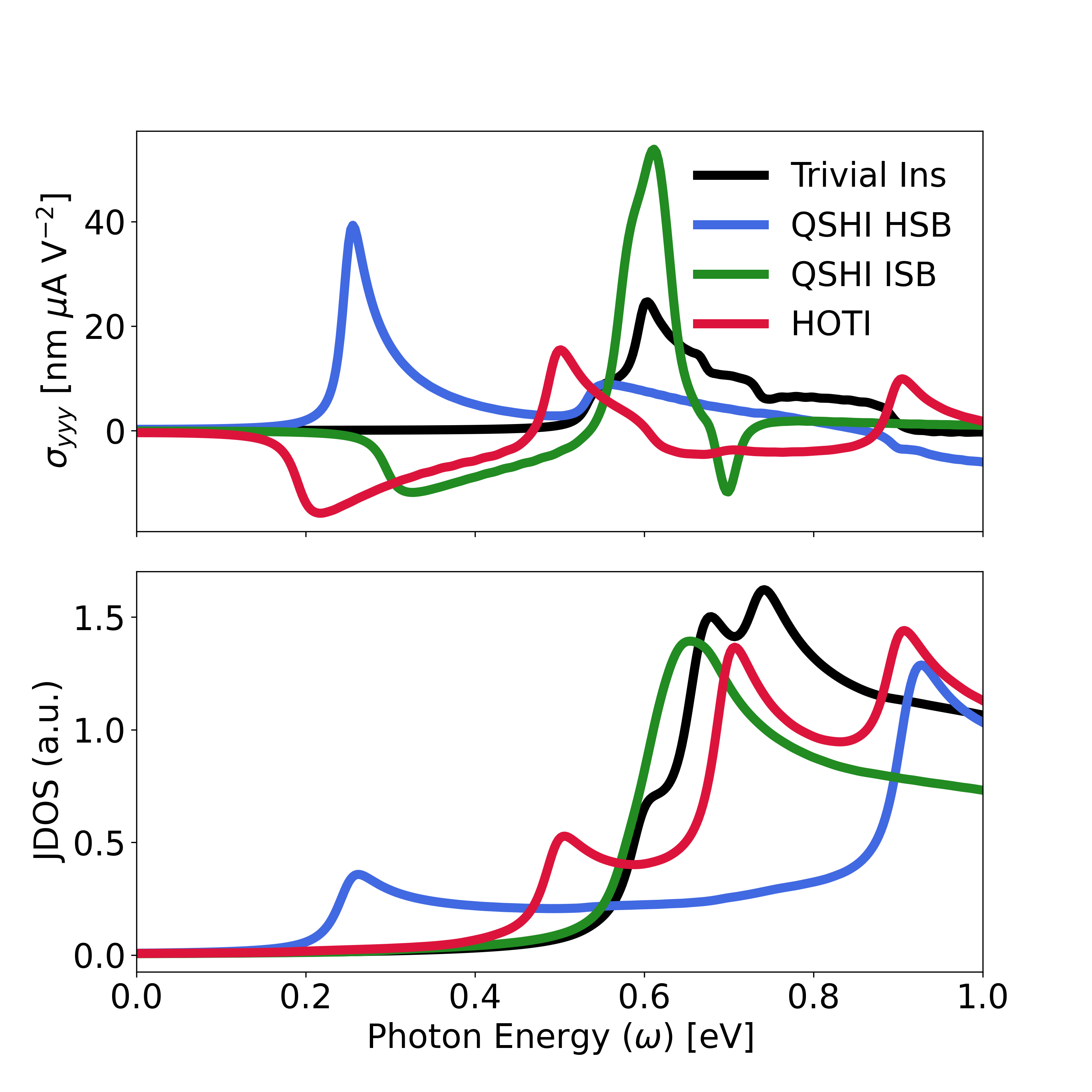}
    \caption{Tensor component $\sigma^{yyy}_{sc}$ (top panel) and joint density of states (bottom panel) of the shift current for the three different topological phases compared with the trivial case.}
    \label{fig:SC-yyy_JDOS}
\end{figure}

To investigate how real-space obstruction influences the nonlinear optical response, we evaluate the SC conductivity of the four insulating phases described by the Hamiltonian in Eq.~(\ref{Eq:TB}). The second-order direct photocurrent generated by monochromatic linearly polarized light is described by the SC conductivity tensor, which has also recently been generalized in terms of the quantum geometric tensor~\citep{Ahn_2021_SC,Verma_2026_SC,Avdoshkin_2025_SC}, in the charge channel~\citep{sipe2000second,IbaezAzpiroz2018_eqsc}

\begin{equation}
\label{eq:sigma_sc}
\begin{aligned}
\sigma_{sc}^{abc}(0; \omega, -\omega) = -\frac{i\pi e^3}{4\hbar^2}
\sum_{\vec{k}} \sum_{n,m} (f_{m\vec{k}}-f_{n\vec{k}}) \times \\
\times \big(r_{mn}^b r_{nm}^{c;a} + r_{mn}^c r_{nm}^{b;a}\big)
\big(\delta(\omega_{mn}-\omega) + \delta(\omega_{nm}+\omega)\big).
\end{aligned}
\end{equation}

where $r_{nm}^{a} = i\langle u_n|\partial_{k_a}|u_m\rangle = A_{nm}^{a}$ denotes the interband dipole matrix element, equivalently given by the interband Berry connection, while $r_{nm}^{c;a}$ is the corresponding generalized derivative. \\
The SC originates from interband optical transitions and is governed by two distinct ingredients: the probability of exciting an electron between occupied and empty states, encoded by 
\begin{equation}
\alpha_{nm}^{ab} = v^a_{nm}v^b_{mn}
\end{equation}
with $v^a_{nm} = i\omega_{nm} r^a_{nm}$ is the interband velocity matrix element between bands $n$ and $m$; and the real-space displacement of the electronic wave packet accompanying the transition. This second contribution becomes explicit once the generalized derivative is introduced

\begin{equation}
r_{nm}^{c;a} = -ir_{nm}^{c}R_{nm}^{ac},
\end{equation}

where $R_{nm}^{ac}$ is the shift vector, measuring the displacement of the electronic charge center during photoexcitation.
In the following, all quantities are evaluated using a customized WannierBerri package~\citep{tsirkin2021high}. The SC response is given by the product of the optical transition probability and the shift vector, highlighting its quantum geometric origin. Since the topological phases discussed in the previous section differ primarily by the arrangement of their Wannier centers, we expect the largest variations of the nonlinear response to originate from modifications of the shift vector rather than from changes in the optical transitions. To verify this picture, we show in Fig.~\ref{fig:SC-yyy_JDOS} the SC conductivity and the JDOS for the four insulating phases listed in Table~\ref{table:TBparameters}. The JDOS, which quantifies the number of available transitions at a given photon energy, is provided by the shape of the electronic structure. In particular, the trivial insulating phase displays a JDOS comparable to, or even larger than, those of the topological phases. In contrast, the highest SC response are enhanced in the QSHI (peak at $\sim 0.6$ eV for ISB and at $\sim 0.25$ eV for HSB) and HOTI (peak at $\sim 0.2$ eV) phases, whereas the trivial phase exhibits the weakest photocurrent. This comparison demonstrates that the enhanced nonlinear response is dominantly given by the geometrical contribution encoded in the shift vector, which is dramatically modified by the real-space obstruction of the wavefunction.\\

\section{Application to the quantum spin Hall insulator Indenene}

\begin{figure*}
    \centering
    \includegraphics[width=\linewidth]{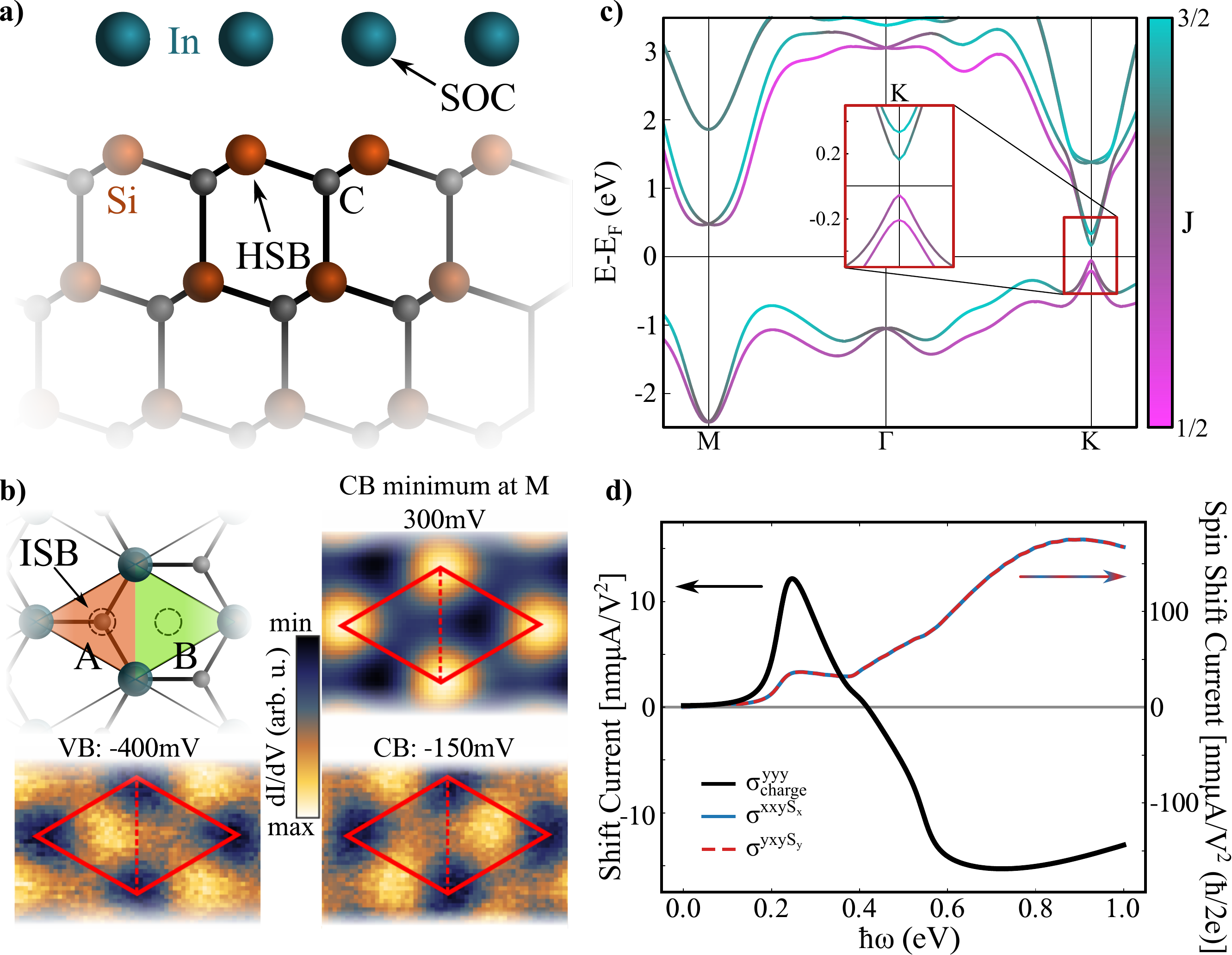}
    \caption{a) and b) represent the top and side view of the crystal structure of indenene. In b) also the experimental Scanning Tunneling Microscopy topographic images at different bias voltages are shown to illustrate the real-space obstruction of indenene, with the conduction and valence bands possessing opposite charge localization within the unit cell. c) Computed band structure of indenene fitted from first principle calculations with total angular momentum projection represented by the colormap. The inset show a zoom in of the Dirac cone around the $K$ point. d) Calculated components of the charge (black line referred to the left y-axis) and spin (colored line referred to the right y-axis) shift-like currents of indenene. Figure reprinted and adapted from Ref.~\citep{Bauernfeind_2021_indenene_topology} under a Creative Commons Attribution 4.0 International License (CC BY 4.0).}
    \label{fig:indenene}
\end{figure*}

The minimal model introduced above captures the essential ingredients realized in a broad class of surface materials consisting of a monolayer of (p)-shell adatoms deposited on group-IV substrates, such as SiC(0001) or Si(111). In these systems, the substrate naturally provides the ISB and HSB terms, while the SOC of the adatoms completes the ingredients required to realize the different topological phases discussed above. In the following we propose a monolayer of indium on SiC(0001), commonly referred to as indenene (Fig.~\ref{fig:indenene}a,b), which has been identified as a candidate HSB-driven QSHI with obstructed Wannier centers~\citep{Bauernfeind_2021_indenene_topology,Erhardt_2022_indenene_topology,Eck_2022_indenene_topology,Erhardt_2024_indenene_topology,Erhardt_2025_indenene_topology,Schmitt_2024_indenene_topology_QSHI}. To assess whether the mechanism identified in the minimal model persists in a realistic material, we construct the TB model developed in Ref.~\citep{Erhardt_2025_indenene_topology}, which accurately reproduces the electronic structure of indenene. The corresponding band structure is shown in Fig.~\ref{fig:indenene}c and displays the characteristic features of the HSB-driven QSHI phase (Fig.~\ref{fig:structure}c). We then calculate the SC spectrum of indenene, with the results reported in Fig.~\ref{fig:indenene}d. The realistic indenene model exhibits a pronounced nonlinear photocurrent, with a magnitude which is comparable to, or even exceeds, that reported in other materials with large nonlinear optical responses~\citep{Bagaglini_moire}\citep{Xu2021_SSC}.

To further investigate the optical response, we also compute the spin shift-current (SSC) conductivity $\sigma_{\mathrm{sh}}^{abcs_i}$ under circularly polarized light, following the velocity-gauge formalism introduced in Ref.~\citep{Xu2021_SSC}. The response is evaluated using the spin-current operator 

\begin{equation} 
j^{a,s_i} = \frac{1}{2} \left\{ v^a,s_i \right\}, 
\end{equation} 

where $s_i=\frac{\hbar}{2}\sigma_i$ ($i=x,y,z$) is the spin operator and $\{\cdot,\cdot\}$ denotes the anticommutator. The full expression for the spin shift-like conductivity and its implementation are reported in the Supplementary Materials~\citep{SM}. To allow a direct comparison with the charge shift-current conductivity, the spin conductivity is reported in units of $\hbar/(2e)$, following the convention of Ref.~\citep{Xu2021_SSC}. The calculated spin shift-like response is shown by the colored curves in Fig.~\ref{fig:indenene}d. For light in the $xy$ plane, the nonvanishing longitudinal components satisfy \begin{equation} \sigma_{\mathrm{spin}}^{xxys_x} = \sigma_{\mathrm{spin}}^{yxys_y}, \end{equation} while reversing the light indexes exchanges the optical indices and reverses the sign of the response, \begin{equation} \sigma_{\mathrm{sh}}^{axys_i} = -\sigma_{\mathrm{sh}}^{ayxs_i}. \end{equation}
It is worth noting that due to the different symmetry rules of the charge and spin SC tensor it is possible to switch spin or charge currents depending on the polarization of the light.

\section{Conclusions}

In conclusion, we have investigated the nonlinear optical response, namely the shift-current, of a class of QSHI on a triangular lattice that show a complete topological phase diagram controlled by the spin-orbit coupling and the out-of-plane and in-plane inversion-symmetry-breaking terms. We have shown that the shift current can be strongly enhanced in topologically nontrivial phases even when their band structures remain similar. This demonstrates that the real-space geometry of the electronic wavefunctions is the key ingredient for having a strong response. We have further demonstrated that this mechanism extends to a realistic material platform by considering indenene, an HSB-driven obstructed QSHI. Its sizable shift-current response confirms that the enhancement associated with real-space obstruction persists beyond the minimal model. Moreover, we have shown that the distinct symmetry selection rules governing charge and spin currents can lead to a finite spin shift current in the absence of a charge shift current, providing a way for generating a pure spin current through THz optical pumping.
Generally, our results sets real-space obstruction as a possible controllable parameter for engineering nonlinear optical responses in surface materials, such as transition metal dichalcogenides, recently measured to posses an electronic obstruction~\citep{Clugru_2026_Obstructed_TMD,Holbrook_2026_Obstructed_TMD}. Since the degree of real-space obstruction is determined by the hybridization between the adatom layer and the substrate, the shift-current response can potentially be tuned through the choice of adatom and substrate, e. g. Tl/Si(111) which also exhibits a larger optical gap of more than 1 eV~\citep{Stolwijk_2013_Tl_Si,Stolwijk_2015_Tl_Si,Erhardt_2026_Tl_Si}, potentially suitable for photovoltaic applications, as well as by external perturbations such as electric fields or strain. These tunable parameters provide a possible experimental control of the underlying topological phases and their associated nonlinear optical responses, opening opportunities for the experimental exploration of geometry-driven photogalvanic and spin-photocurrent phenomena in surface systems.

\section*{Acknowledgements}

R.C., G.S. and J.E. are grateful for funding support from the Deutsche Forschungsgemeinschaft (DFG, German Research Foundation) under Germany's Excellence Strategy through the W\"urzburg-Dresden Cluster of Excellence on Complexity, Topology and Dynamics in Quantum Matter ctd.qmat (EXC 2147, Project ID 390858490) as well as through the Collaborative Research Center SFB 1170 ToCoTronics (Project ID 258499086). M.I. and G.S. acknowledge funding support from the FOR 5249 [QUAST, Project ID 449872909 (Project P5)],

\bibliography{bibliography}
\end{document}